\documentclass[aps,prl,twocolumn,floatfix,longbibliography, superscriptaddress]{revtex4-1}
\usepackage{graphicx}
\usepackage{amsmath}
\usepackage{amssymb}
\usepackage{physics}
\usepackage{braket}
\usepackage{tikz}
\usepackage{float}
\usepackage[caption = false]{subfig}

\begin{document}

\title{Frustration in a dipolar Bose-Einstein condensate introduced by an optical lattice}
\begin{abstract}
We study the application of a square perturbing lattice to the naturally forming hexagonal arrays of dipolar droplets in a dipolar Bose-Einstein condensate. We find that the application of the lattice causes spontaneous pattern formation and
leads to frustration in some regimes. For certain parameters, the ground state has neither the symmetry of the intrinsic hexagonal supersolid nor the symmetry of the square lattice. These results may give another axis on which to explore dipolar Bose-Einstein condensates and to probe the nature of supersolidity. 
\end{abstract}

\author{Eli J. Halperin}
\affiliation{JILA, NIST, and Department of Physics, University of Colorado, Boulder, Colorado 80309-0440, USA}
\author{Shai Ronen}
\affiliation{Cleerly Inc., 110 16$^{th}$ St., Suite 1400, Denver, CO, 80218, USA}
\author{J. L. Bohn}
\affiliation{JILA, NIST, and Department of Physics, University of Colorado, Boulder, Colorado 80309-0440, USA}
\date{\today}
\maketitle

Geometric frustration is important throughout nature, and is particularly relevant to the study of many-body interacting systems such as spin liquids and spin glasses. One of the hallmarks of frustration is the inability of the system to find a unique and fully periodic ground state configuration~\cite{ramirez2003geometric}. Frustration occurs when conflicting interactions between discrete constituents each favor some specific configuration, and yet no regular configuration is found. A famous example is antiferromagnetic order on a triangular lattice, where pairs of neighboring spins cannot all be anti-aligned simultaneously~\cite{PhysRev.79.357}. For frustrated Ising spins on a square lattice~\cite{vannimenus1977theory}, ferromagnetism suddenly vanishes below a certain interaction strength. Part of the ongoing interest in quantum simulation is to generate frustrated systems under carefully controlled conditions, such as in ion crystals~\cite{islam2013emergence} or optical lattices~\cite{struck2011quantum}, which mimic the properties of frustrated magnetic systems, in order to glean some of their elusive properties. 

Here we consider a system that is already self-organizing in interesting ways, namely, a Bose-Einstein condensate (BEC) whose constituent atoms are magnetic and hence interact via dipolar interactions, as is relevant in Dysprosium~\cite{lu2011strongly}, Erbium~\cite{aikawa2012bose}, or Chromium~\cite{griesmaier2005bose}.  Such a BEC can be coaxed into a supersolid state~\cite{ferliano2019supersolid, bisset2015crystallization, tanzi2019observation, norcia2021two}, exhibiting periodic ordering while retaining the coherence properties of the superfluid.  The observation of this state was a major experimental milestone~\cite{ferliano2019supersolid, norcia2021two, PhysRevX.9.011051, ferrier2016observation}, realizing predictions going back to speculations in superfluid helium~\cite{PhysRevA.2.256}. Under other circumstances, other novel configurations of density are predicted~\cite{hertkorn2021pattern, halder2022control}.

\begin{figure}
    \centering
    \includegraphics[width = 0.48\textwidth]{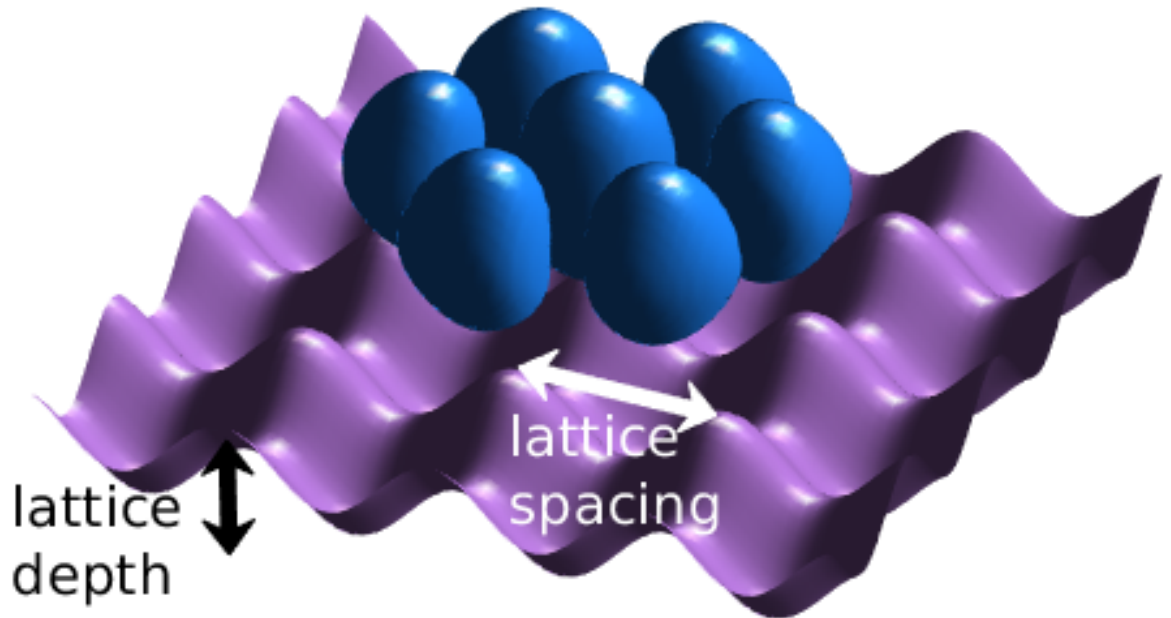}
    \caption{A schematic illustration of the unperturbed droplet ground state, shown by blue isodensity surfaces, with the applied perturbing lattice, shown in purple. The lattice has an incommensurate symmetry with the natural ground state, and thus the droplets struggle to simultaneously fulfill the constraints of their interactions and of the lattice.}
    \label{fig:schematic}
\end{figure}

We return to the supersolid state for inspiration, noting that not all solids are crystalline in nature.  Density modulations may be aperiodic, as in a glassy state; or they may have spatially distinct regions of differing symmetries, indicating frustration.  Here we place uncomfortable stress of the DBEC's native sixfold structure, by subjecting it to an optical lattice of fourfold symmetry, as shown schematically in Fig.~\ref{fig:schematic}. In the process, a rich variety of states occurs, including some that are frustrated by this competition.

In typical treatments of DBECs, such as the approach we apply here, one considers only a single order parameter to represent the many-body wavefunction, where each atom in the DBEC has the same wavefunction. The long-range phase coherence between spatially distinct droplets has been experimentally verified in 1 dimensional systems~\cite{ferliano2019supersolid}, and is seen here numerically in 2 dimensional states. Frustration, on the other hand, relies on distinct components having competing interactions with one another, where no such phase coherence is required. Here, we describe a distinctly quantum mechanical version of frustration in a single coherent field, where different spatial regions of the field organize according to different governing principles.

The depth of individual lattice sites and the spacing between them become essential for determining the ground state. Frustration may occur due to the difference, or competition, between the optimal geometry of the system without the lattice, and the geometry of the lattice. For certain lattice depths and spacings, the ground state energetic manifold may become highly degenerate due to the application of the lattice. We examine ground state morphologies as a function of the lattice spacing and lattice depth. We map out several different phases in this regime, where, in particular, the density forms checkerboard, stripe, and frustrated patterns.  

\emph{Model.---}
DBECs can be described by the extended Gross-Pitaevskii equation (EGPE)~\cite{dalfovo1999theory}, where the ground state condensate order parameter $\psi$ obeys
\begin{align}
&\mu \psi(\vec r) = \bigg[-\frac{\hbar^2}{2m} \nabla + V + g\vert\psi(\vec r) \vert^2 \nonumber \\
&+ \int U_{dd}(\vec r - \vec r') |\psi(\vec r')|^2 + \gamma_{QF} |\psi(\vec r)|^3 \bigg] \psi(\vec r).
\label{GPE}
\end{align}
Here $\mu$ is the chemical potential and $V = V_{ext} + V_{lat}$ the external potential, which may include both a harmonic trap $V_\text{ext}$ and an applied lattice $V_\text{lat}$
\begin{align}
    V_{\text{ext}} &= \frac{1}{2} m  (\omega_x^2 x^2 + \omega_y^2 y^2 + \omega_z^2 z^2), \\
    V_{\text{lat}} &= \frac{V_0}{4} \left( \cos k x + \cos k y \right),
\end{align}
where $\omega_{x,y,z}$ gives the trap frequency along the corresponding direction. $2 \pi /k$ gives the lattice spacing and $V_0$ the peak-to-trough lattice depth. In the above equations, $g = 4 \pi \hbar^2 a/ m$ is the contact interaction strength, with $a$ the s-wave scattering length. $U_{dd}$ accounts for the long range dipole-dipole interaction, and is given by
\begin{align}
    U_{dd}(\vec r) = \frac{3 \hbar^2 a_{dd}}{m r^3} \left( 1 - 3 \cos^2(\theta) \right),
\end{align}
where $a_{dd} = m\mu_0\mu^2/12\pi\hbar^2$ is the dipole length, $r$ the distance between two interacting dipoles, and $\theta$ the angle between $\vec r$ and the dipole alignment axis, here taken to be $\hat z$. $\gamma_{QF}$ arises from the local-density approximation to quantum fluctuations~\cite{lima2011quantum, baillie2016self, LHY1, LHY2}, and is 
\begin{align}
    \gamma_{QF} = \frac{32}{3} g \sqrt{\frac{a^3}{\pi}} \left(1 + \frac{3 a_{dd}^2}{2 a^2} \right).
\end{align}
For time-dependent calculations, we simply replace $\mu \psi$ by $i \hbar~\partial \psi / \partial t$ in Eq.~\eqref{GPE}. The EGPE energy functional is
\begin{align}
    &E[\psi] = \int \bigg[ \frac{\hbar^2}{2m} |\nabla \psi|^2 + V |\psi_0|^2 + \frac{1}{2} g\vert\psi(\vec r) \vert^4 \nonumber \\
&+ \frac{1}{2} \int U_{dd}(\vec r - \vec r') |\psi(\vec r')|^2 |\psi(\vec r)|^2 d\vec r'  + \frac{2}{5} \gamma_{QF} |\psi(\vec r)|^5 \bigg] d \vec r.
\label{energy_func}
\end{align}
We numerically minimize this functional using a limited-memory Broyden–Fletcher–Goldfarb–Shanno (LBFGS) algorithm~\cite{liu1989limited} optimized on a graphics processing unit (GPU). We use PyTorch's LBFGS algorithm taking advantage of automatic differentiation. We solve for the ground state on a cubic grid of $128 \times 128 \times 96$ grid points of size $28 \times 28 \times 28 \,~a_{ho}$, with $a_{ho}$ the oscillator length. A large enough grid is used along with a cut-off in the maximum size of the dipolar interaction, i.e. $U_{dd}(\vec r) = 0$ for $|\vec r| > 14~a_{ho}$. As long as the grid as twice is large as the cutoff, and the ground state density all lies within a sphere of radius $14~a_{ho}$, then this ensures that ``phantom condensates'' are not considered~\cite{Ronen2007Roton} while introducing no other effects. 

Due to the elaborate energy landscape, there are sometimes many metastable states where the energy functional has zero gradient. Moreover, some of these metastable states appear highly attractive to initial conditions with added Perlin~\cite{perlin1985image, hertkorn2021pattern} and uniform noise, where the minimization procedure can find a metastable state instead of the true ground state. Such states are verified to be metastable by performing time-evolution via a time-splitting pseudospectral method~\cite{bao2012mathematical}. In order to address the complicated energy landscapes, we first perform a minimization with over $50$ random initial conditions for each set of parameters. Then, we identify the prominent morphologies seen over all lattice depths and spacings. The wavefunctions of these distinct morpohlogies are then used to re-seed the minimization at all of the considered sets of parameters. Procedures which do not follow such an iterative method may misidentify some phase boundaries. 
Henceforth, we consider $^{162}$Dy atoms in a $\left( \omega_x,\omega_y,\omega_z \right) / 2 \pi = 125 \times 125 \times 250$ Hz trap. The dipole length of $^{162}$Dy is $a_{dd} = 131 a_0$ and we use a scattering length $a = 85 a_0$. The atom number crucially determines the density profiles of the DBEC, and we choose the total atom numbers of $N = 10^5$. For this system, $a_{ho} = \sqrt{\hbar / m \omega} = 0.71~\mu m$ and $\hbar \omega_x / k_B = 6~nK$.

\begin{figure*}[htb]
    \centering
    \includegraphics[width = \textwidth]{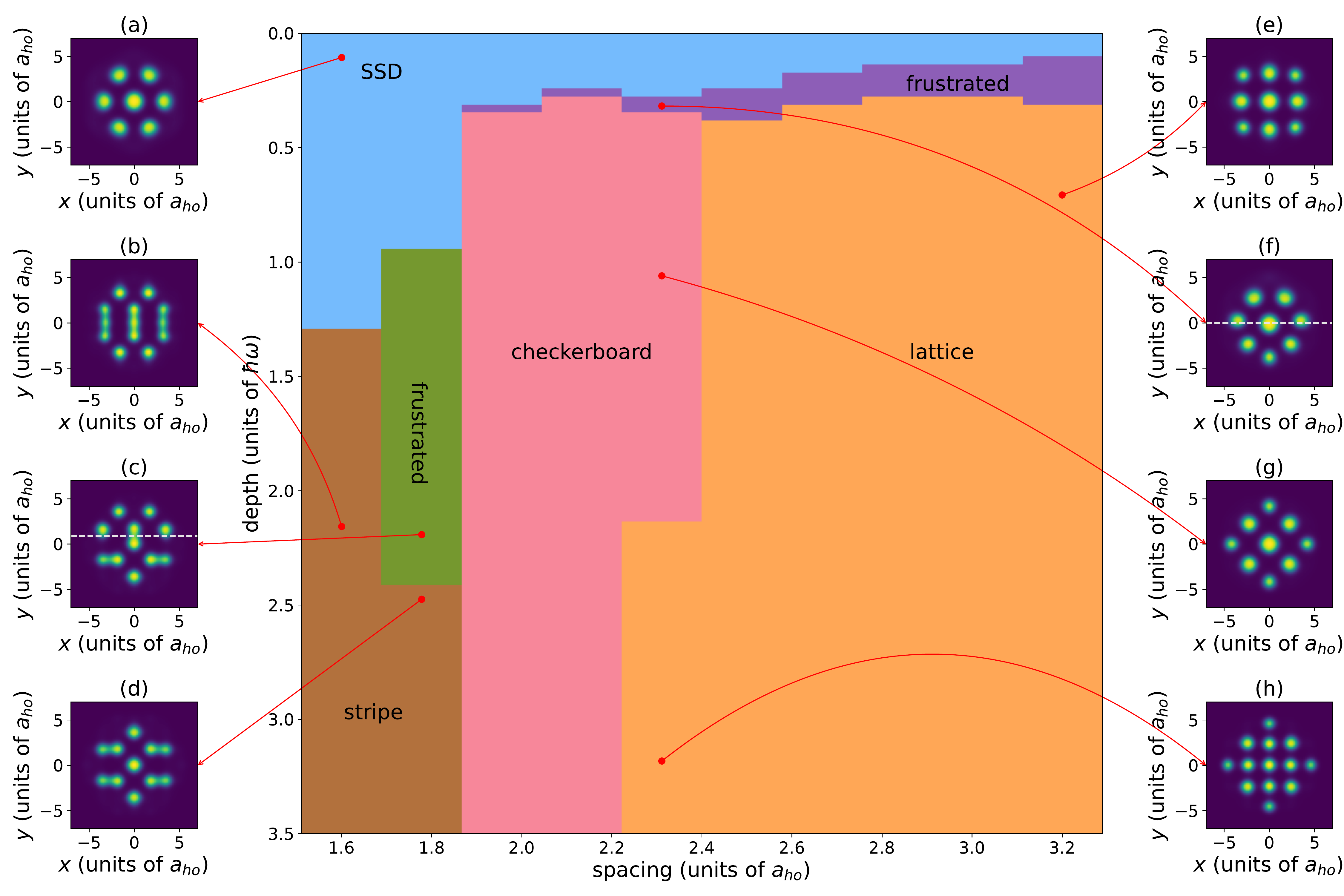}
        \caption{The supersolid droplet phase for $N=100000$ $^{162}$Dy atoms in $125 \times 125 \times 250$ Hz trap, when a square perturbing lattice potential is applied. Sample phases are shown on the left and right, showing density slices in the $xy$ plane through the center of the trap. The lattice parameters used to generate these phases given by the tail of the red arrows, and the patterns formed are colored accordingly.}
        \label{fig:ssd_square}
\end{figure*}

\emph{Lattice.---}
We consider the effect of applying a weak lattice with 4-fold symmetry to the supersolid ground state. Figure~\ref{fig:ssd_square} shows the resulting phase diagram as well as select states that occur under the conditions indicated by the red lines. This diagram is based on a grid with 100 values of lattice depth from top to bottom, and 10 values of lattice spacing from left to right. Within this resolution, the different colored regions represent different phases of organization of the DBEC, which are labeled to indicate their general character. On left and right of this figure, selected density profile along 2D slices through the center of the trap are shown. The central figure shows the four distinct phases this state as the lattice depth and spacings are varied.

Along the top of the diagram, the blue section labelled SSD shows the parameters for which the DBEC is in a 6-fold symmetric state, closely resembling the unperturbed ground state. This occurs when the lattice is fairly weak, or when the lattice is so tightly spaced the the energetic cost of dipolar droplets sitting in the lattice minima would be quite large, as is true for a lattice spacing of $1.6~a_{ho}$ up to a depth of 1.31~$\hbar \omega$. An unperturbed SSD density profile is shown in Fig.~\ref{fig:ssd_square}(a).

In the opposite limit, of deep lattices (lower right corner), the lattice dominates the physics. When the lattice spacing becomes comparable to the intrinsic spacing of $3.2 a_{ho}$, the DBEC transitions abruptly to the 4-fold symmetric state with each lattice minima occupied by a dipolar droplet (lattice, orange). Density slices of two such states are shown in Fig.~\ref{fig:ssd_square}(e) and Fig.~\ref{fig:ssd_square}(h), the first at a lattice spacing of $3.2~a_{ho}$ and depth of $0.71~\hbar\omega$, and the second at a lattice spacing of $2.31~a_{ho}$ and depth of $3.18~\hbar\omega$. Here, either 9 or 13 droplets form. Similar states were seen to be metastable in the absences of an lattice potential~\cite{young2022supersolid}. The SSD and lattice phases indicate no frustration; the density profiles entirely satisfy one of the constraints placed on them. Between these regimes, the DBEC cannot ignore the lattice, nor is it completely in thrall to it.

In this intermediate regime, one find a class of states that have a checkerboard density pattern, prominently featured in the diagram (checkerboard, pink). Here, the lattice spacing is somewhat smaller than the SSD spacing of around $3.2~a_{ho}$. Thus, the DBEC can maintain a larger spacing between droplets by simply filling every-other lattice site while partially satisfying the constraints of the lattice. An example checkerboard density profile is shown in Fig.~\ref{fig:ssd_square}(g), for a lattice depth of $1.06~\hbar\omega$ and spacing of $2.13~\hbar \omega$. Here, there are four unoccupied lattice sites within the trap, which for comparison are seen to be filled in Fig.~\ref{fig:ssd_square}(h), at the same the spacing but a depth of $3.18~\hbar \omega$. Here, the repulsion between droplets is strong enough at this short-range that the energetic cost of occupying adjacent lattice sites would be too high.
The checkerboard state belies a competition between the interatomic forces and the applied lattice, however there are not distinct regions of space with different density patterns, as one might expect from a frustrated system. 

SSD, lattice, and checkerboard phases all retain some symmetry, but in the regime between these, anomalous phases appear. As the lattice depth increases from $0.2~\hbar \omega$  to $0.3~\hbar \omega$ at spacings around $2~a_{ho}$ the DBEC transitions between the SSD and checkerboard state. In this regime, the DBEC becomes frustrated, forming the density pattern shown in Fig.~\ref{fig:ssd_square}(f). In the top half of this density profile, the density closely resembles that of the SSD state, as shown in Fig.~\ref{fig:ssd_square}(a), while in the bottom half of this density profile, the density resembles the checkerboard, as in Fig.~\ref{fig:ssd_square}(g). The energy of this frustrated state is lower than either the checkerboard or the SSD at this lattice depth. Here, the inter-droplet repulsion is so strong that several lattice sites being empty between droplets may be energetically favorable.
Different spatial regions form distinct density patterns, one preferred by the SSD morphology and the other by the checkerboard. Typical densities between droplets is $0.1 - 10\%$ of the peak density, which acts as evidence of the phase coherence in these systems.

\begin{figure}
    \centering
    \includegraphics[width = 0.48\textwidth]{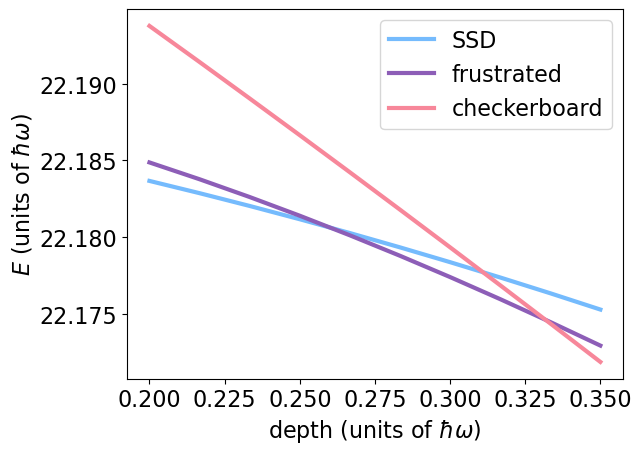}
    \caption{The energies as a function of increasing lattice depth, at the fixed lattice spacing of $2.31~a_{ho}$, with the minimization seeding by the three dominant morphologies in that region: the SSD, the frustrated state, and the checkerboard state.}
    \label{fig:transition}
\end{figure}

The detailed transition from the SSD to checkerboard phase is traced in the high-resolution slice of the data in Fig.~\ref{fig:transition}, for fixed lattice spacing of $2.31~a_{ho}$. This includes states similar to the three shown in Figs.~\ref{fig:ssd_square}(a), \ref{fig:ssd_square}(f), and \ref{fig:ssd_square}(g). By seeding, the minimization can be forced to find any of the three shown states, one of which is always the ground state. Thus, we can extract the energy of these three states, where the details of the state are re-optimized for the specific system parameters, yet the overall morphology is unchanged. We indeed find a region where the frustrated state is the lowest energy, with the SSD being lower energy at smaller depths, and the checkerboard being lower energy at large depths. Between these different regimes, we find discontinuities in the derivative of the ground state energies, indicating second-order phase transitions. Such phase transitions as a function of lattice depth appear through the diagram. 

At smaller lattice spacing (left side of Fig.~\ref{fig:ssd_square}), the DBEC forms states with one or more stripes (brown, stripe), that again have neither $4-$ nor $6-$fold symmetry. Stripe states were also seen at slightly higher atom number in an unperturbed DBEC~\cite{hertkorn2021pattern}. Two such states are shown in Fig.~\ref{fig:ssd_square}(b) and Fig.~\ref{fig:ssd_square}(d), where there are large gaps between sometimes connected filled lattice sites.

Between the stripe and checkerboard regime, we find another frustrated ground state, indicated by the green region in Fig.~\ref{fig:ssd_square}. This frustrated state is shown in Fig.~\ref{fig:ssd_square}(c), where the bottom portion of the density profile, below the dotted white line, closely matches that of Fig.~\ref{fig:ssd_square}(d), while the top portion, above the dotted white line, matches the density of both a checkerboard state at this smaller lattice spacing and the stripe phase seen in Fig.~\ref{fig:ssd_square}(b), with alternating filled lattice minima. Here, the boundary between these two morphologies does not pass through the center of the trap, but instead is slightly above it. Given that we seed the entire phase diagram with all stripe states, this frustrated intermediate state is lower energy than either of the two halves which comprise it. 

\begin{figure}
    \centering
    \includegraphics[width = 0.48\textwidth]{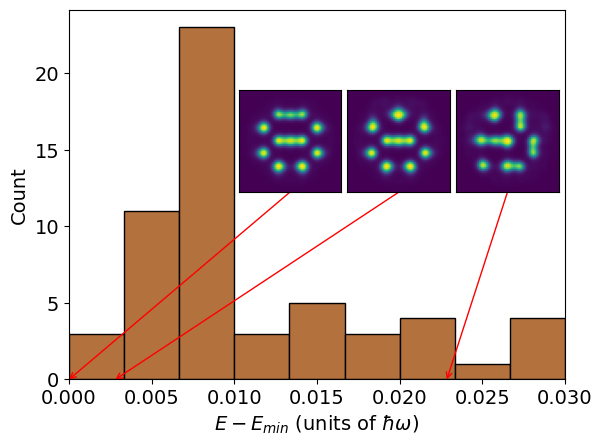}
    \caption{A histogram of different energies found by the minimization procedure showing a subset of 100 different initial conditions, at a lattice depth of $1.94~\hbar \omega$ and spacing of $1.6~a_{ho}$. Three example densities are shown as insets, with the corresponding energies indicated by red arrows.}
    \label{fig:glassy}
\end{figure}

The stripe regime is emblematic of the complicated energy manifold that arises when the lattice is applied. In the SSD regime, without a lattice, the ground-state is energetically well-separated from metastable states. However, in the stripe regime, there are a plethora of metastable states with small energy spacings from the ground state. This is shown in Fig.~\ref{fig:glassy}, where a histogram of energies found by the minimization is shown, along with three example states, at a lattice spacing of $1.6~a_{ho}$ and depth of $1.94~\hbar \omega$. Here, the ground state finds a different stripe phase than shown in either Fig.~\ref{fig:ssd_square}(b) or Fig.~\ref{fig:ssd_square}(d). The nearest distinct state has a spacing of $7 \times 10^{-4}~\hbar \omega$ from the ground state. In some cases, nearby states are similar to the ground state, while in others they are entirely unrelated, as shown by the inset density profiles. The shape of the histogram belies that the lowest energy state may not always be the most energetically attractive local minimum to a random initial state, and in this case the ground state is not the most likely state for the minimizer to find. 

\emph{Outlook.---}
We have investigated the effect of a perturbing lattice on the supersolid ground state of a DBEC. In some cases the DBEC has neither the symmetry of its natural ground state nor the symmetry of the lattice.  Undiscovered states may yet exist outside of the boundaries of our phase diagram, between points in the scan, or in isolated regions of the energy landscape. This sets off the exploration of a vast phase space, where different lattices or other perturbing potentials could be applied to the self-organizing patterns of a DBEC. One could begin from the labyrinthine superglass or honeycomb pattern~\cite{hertkorn2021pattern}, and consider the effects of a lattice on that system. The application of a time-dependent lattice, either adiabatically or via a quench, additionally remains unexplored. 

This material is based upon work supported by the National Science Foundation under Grant Number PHY 1734006.

\bibliography{dipoles_lattice}

\end{document}